\documentclass[conference]{IEEEtran}
\IEEEoverridecommandlockouts
\usepackage{cite}
\usepackage{amsmath,amssymb,amsfonts}
\usepackage{algorithmic}
\usepackage{adjustbox}
\usepackage{graphicx}
\usepackage[para,online,flushleft]{threeparttable}
\usepackage{textcomp}
\usepackage{multirow}
\usepackage[para,online,flushleft]{threeparttable}
\usepackage[dvipsnames]{xcolor}
\def\BibTeX{{\rm B\kern-.05em{\sc i\kern-.025em b}\kern-.08em
    T\kern-.1667em\lower.7ex\hbox{E}\kern-.125emX}}
\begin{document}

\bstctlcite{BSTcontrol}

\title{Pushing up to the Limit of Memory Bandwidth and Capacity Utilization for Efficient LLM Decoding on Embedded FPGA}

\author{Jindong Li$^{1, 2, 4}$ \ \ Tenglong Li$^{1, 2, 4}$ \ \ Guobin Shen$^{1, 2, 3}$ \ \ Dongcheng Zhao$^{1,2}$ \ \  Qian Zhang$^{1, 2, 4}$ \ \ Yi Zeng$^{1, 2, 3, 4, 5}$\\
$^1$Brain-inspired Cognitive Intelligence Lab, Institute of Automation, Chinese Academy of Sciences\\ 
$^2$ Center for Long-term Artificial Intelligence \\
$^3$ School of Future Technology, $^4$ School of Artificial Intelligence, University of Chinese Academy of Sciences  \\ 
$^5$ Key Laboratory of Brain Cognition and Brain-inspired Intelligence Technology, Chinese Academy of Sciences\\ 
{\{lijindong2022, litenglong2023, shenguobin2021,}{zhaodongcheng2016, q.zhang, yi.zeng\}@ia.ac.cn}
\thanks{Corresponding author: q.zhang@ia.ac.cn and yi.zeng@ia.ac.cn.}
}

\maketitle

\begin{abstract}
The extremely high computational and storage demands of large language models have excluded most edge devices, which were widely used for efficient machine learning, from being viable options.
A typical edge device usually only has 4GB of memory capacity and a bandwidth of less than 20GB/s, while a large language model quantized to 4-bit precision with 7B parameters already requires 3.5GB of capacity, and its decoding process is purely bandwidth-bound.
In this paper, we aim to explore these limits by proposing a hardware accelerator for large language model (LLM) inference on the Zynq-based KV260 platform, equipped with 4GB of 64-bit 2400Mbps DDR4 memory.
We successfully deploy a LLaMA2-7B model, achieving a decoding speed of around 5 token/s, utilizing 93.3\% of the memory capacity and reaching 85\% decoding speed of the theoretical memory bandwidth limit.
To fully reserve the memory capacity for model weights and key-value cache, we develop the system in a bare-metal environment without an operating system. To fully reserve the bandwidth for model weight transfers, we implement a customized dataflow with an operator fusion pipeline and propose a data arrangement format that can maximize the data transaction efficiency.
This research marks the first attempt to deploy a 7B level LLM on a standalone embedded field programmable gate array (FPGA) device. It provides key insights into efficient LLM inference on embedded FPGA devices and provides guidelines for future architecture design.

\end{abstract}


\section{Introduction}

Large Language Models (LLMs) have revolutionized natural language processing (NLP) by enabling applications ranging from chatbots to complex analytical tools. These models can generate human-like text, provide insightful analysis and assist in complex decision-making processes, transforming industries and enhancing user experiences. However, their utility comes at a great cost: LLMs are compute and memory-intensive, requiring substantial resources for training and inference. Yet, cloud-based solution introduces dependencies on internet connectivity and potential concerns about privacy and security.

In the past decade, there has been extensive research on Field Programmable Gate Array(FPGA)-based accelerators for efficient deep learning models like convolutional neural networks\cite{guo2017angel}\cite{chen2021hardware}, spiking neural networks\cite{li2024firefly}\cite{li2024fireflys} and vision transformers\cite{li2022auto}\cite{dong2023heatvit}.
However, when faced with the significantly larger scale of LLM workloads, FPGAs have yet to fully realize their potential.
Recent work, DFX \cite{hong2022dfx} and FlightLLM \cite{zeng2024flightllm}, has demonstrated the feasibility of deploying LLMs on cloud-based Alveo U280 FPGAs with high-bandwidth memory (HBM). However, these approaches are not applicable to embedded FPGAs.
\textbf{The biggest challenge in deploying LLMs onto embedded FPGAs lies in the limited memory resources.} A typical embedded Xilinx FPGA board, such as the Ultra96v2, KV260 or ZCU104, has only 2-4GB DDR4 memory with a speed grade ranging from 1600-2400Mbps, resulting in a total bandwidth of less than 19.2GB/s. LLaMA2-7B \cite{touvron2023llama}, an entry-level LLM, requires around 3.5GB of memory just for the model weights after 4-bit quantization. This does not account for the key-value cache, which grows as the context increases, further escalating the challenge.

To the best of our knowledge, no prior research has attempted to deploy LLMs with 7B parameters on embedded FPGA devices. Therefore, in this paper, we take on this challenge and aim to push the bandwidth and capacity limits of embedded FPGAs for LLM inference. The key contributions of this paper are listed below.

\begin{figure}
    \centering
    \includegraphics[width=1.0\linewidth]{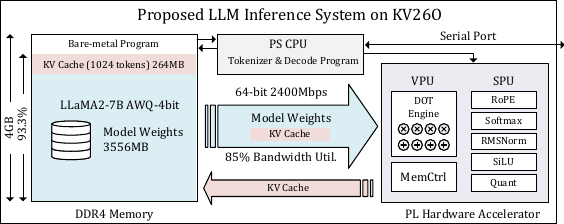}
    \caption{LLaMA2-7B inference on embedded KV260 platform, with 93.3\% of the memory capacity occupied and 85\% of the memory bandwidth utilization.}
    \label{fig:ddrtohardware}
    \end{figure}

1) For the first time, we successfully deployed a 7B parameter LLM on an embedded FPGA, the KV260. With only 4GB of available memory, we developed the system in a bare-metal environment, reserving 93.3\% of the space for model weights and the key-value cache, supporting a context length of up to 1024 tokens, as shown in Fig.\ref{fig:ddrtohardware}.

2) To improve bandwidth utilization for the bandwidth-bound LLM decoding process, we implemented a customized dataflow with an operator fusion pipeline and proposed a custom data arrangement format to maximize data transaction efficiency. Despite the limited 19.2GB/s bandwidth of the KV260 platform, we achieve a decoding speed of around 5 token/s, 85\% of the theoretical decoding limit.

3) We implemented a bandwidth-area optimized hardware architecture that covers all the necessary operations for LLM inference, including LayerNorm\cite{zhang2019root}, Softmax\cite{milakov2018online}, RoPE\cite{su2024roformer}, SiLU\cite{elfwing2018sigmoid}, online quantization, and dequantization.
We successfully fit the architecture within the KV260's limited hardware resource budget and achieved operation at 300 MHz, even with up to 70\% system LUT utilization.

\section{Related Work}

LLMs are based on the Transformer architecture \cite{vaswani2017attention}. 
Hardware architecture for Transformer have been extensively studied in the pre-LLM era, with numerous hardware accelerators optimized for models such as ViT\cite{dosovitskiy2020image} and BERT\cite{devlin2018bert}.
HeatVit\cite{dong2023heatvit} proposed a hardware-efficient image-adaptive token pruning framework for efficient yet accurate ViT acceleration on embedded FPGAs. Liu et al. \cite{liu2021hardware} proposed to fully quantize the BERT and propose an accelerator on Xilinx ZCU102 and ZCU11 FPGA and achieve high power efficiency.
While these methods have proven effective for ViT and BERT models, they cannot be directly applied to LLM workloads.

In the era of LLMs, research on hardware accelerators tailored for LLMs have gradually advanced.
As LLMs are memory-intensive, recent work has focused on accelerating LLMs using HBM-equipped cloud FPGAs, such as the Xilinx Alveo U280 and VCU128, achieving decoding speed improvements over GPUs in single-batch LLM inference setups.
DFX \cite{hong2022dfx} represents one of the first studies on FPGA accelerators for decoder-only transformer, utilizing four banks of Alveo U280 to accelerate the GPT-2 1.5B model \cite{radford2019language}. FlightLLM \cite{zeng2024flightllm} has deployed the LLaMA2-7B model \cite{touvron2023llama} on both the Alveo U280 and the Versal VHK158, demonstrating advantages over NVIDIA GPUs in single-batch inference tasks. Chen et al. \cite{chen2023understanding} provide a comprehensive analytical framework, offering the first in-depth examination of both the benefits and limitations of FPGA-based LLM spatial acceleration using the Alveo U280. EdgeLLM \cite{huang2024edgellm} adopts the VCU128, which is also a very large FPGA equipped with HBM, to deploy LLaMA2-7B, achieving a decoding speed higher than FlightLLM.

While these studies represent the initial steps in FPGA-based LLM acceleration for single-batch decoding, they are not applicable to real-world scenarios. Expensive server-class FPGAs are expected to handle multiple concurrent user inquiries, whereas single-batch decoding is more commonly seen in edge environments.
Accelerating single-batch decoding for LLMs with embedded FPGAs remains a challenging yet unexplored area, which will be the focus of this paper.

\section{LLM Basics}

\begin{figure}
\centering
\includegraphics[width=1.0\linewidth]{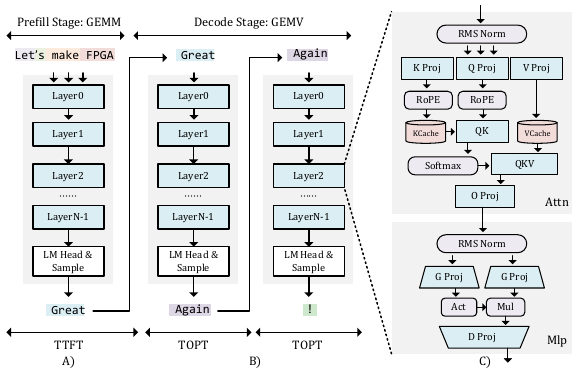}
\caption{LLM Inference Process of a LLaMA-like model. A) The prefill phase. B) The decode phase. C) Inference process breakdown of a single layer.}
\label{fig:llminference}
\end{figure}

A typical LLM model, such as LLaMA, consists of dozens of identical cascading building blocks, with each block comprising an attention layer and an MLP layer, as shown in Fig.\ref{fig:llminference}C.
The attention layer consists of three projection layers to generate the query, key, and value, and the multi-head attention mechanism is applied to the current QKV and the history KV cache. The MLP layer includes up and down projection layers, with an additional gate projection layer applied to the up projection output.

The generative inference process of LLMs consists of two phases: the prefill phase and the decode phase, as shown in Fig.\ref{fig:llminference}. During the prefill phase, the model processes user prompts through the entire model architecture to produce the initial token, as shown in Fig.\ref{fig:llminference}A.
This stage involves processing multiple tokens simultaneously, necessitating compute-bound matrix-matrix multiplication in the linear projection layer.
In the decode phase, the model uses the previously generated token to sequentially generate new tokens in an auto-regressive manner, as shown in Fig.\ref{fig:llminference}B.

In this work, we follow the insights from Chen's analysis\cite{chen2023understanding}: existing FPGAs are less efficient than GPUs during the compute-intensive prefill stage but can outperform GPUs in the memory-intensive decode stage. We focus on optimizing the decode stage of LLM inference.

\section{Algorithmic Optimization}

\subsection{Model Weight Quantization: W4A16}

Post-training quantization (PTQ) has become a common practice to lowers the computational and memory demands of LLMs, as quantization aware training (QAT) of LLM becomes computationally impratical.
FlightLLM employed SmoothQuant \cite{xiao2023smoothquant} to quantize both weights and activations to 8-bit. Later research, AWQ \cite{lin2024awq}, indicated that quantizing weights to a lower bit-width can yield greater speed improvements than quantizing activations in LLM inference.
AWQ quantized weights to 4-bit while keeping activations as 16-bit floating points, achieving less performance loss than SmoothQuant while significantly enhancing decoding performance and reducing memory requirements.
We adopt the W4A16 quantization strategy by AWQ in this work.

\subsection{Key-Value Cache Quantization: KV8}

Quantizing the KV cache can also reduce memory requirements as the context size increases. While it is possible to aggressively quantize the KV cache to 4-bit, it is recommended that for smaller models ($\leq $13B), KV8 quantization is more suitable for preserving capabilities such as multi-step reasoning and self-calibration\cite{li2024evaluating}. In this work, we adopt the KV8 linear quantization strategy, where the quantization process is performed on-chip as the key and value are generated and then sent back to external memory. The dequantization process occurs when fetching the key and value from memory.

\section{Hardware Acceleration Strategy}

\subsection{Dataflow in the Attention Layer}

\begin{figure}
    \centering
    \includegraphics[width=1.0\linewidth]{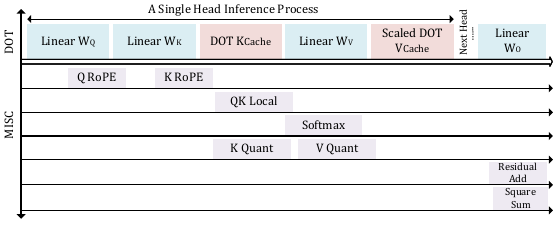}
    \caption{The pipelining dataflow in the attention layer, with all the miscellaneous process hidden in the dense computation to avoid cycle penalties.}
    \label{fig:fusion}
    \end{figure}

\textbf{To achieve high bandwidth utilization, it is crucial to conceal miscellaneous processes during the dense computation to ensure no cycle penalties.}
The attention layer in the transformer inference process involves the majority of miscellaneous operations, making it the focus of our research.

Fig.\ref{fig:fusion} shows the fine-grain pipelining dataflow in the attention layer during the inference process.
Unlike DFX's\cite{hong2022dfx} coarse-grained pipeline where the query, key, and value projections occur before the multi-head attention, we adopt a fine-grained, head-wise pipeline that fuses the projection and attention computation processes.
In the pipeline of each single head, the query projection occurs first, followed by the key projection. During the DOT operation, the query and key are generated element by element. RoPE is applied to the query and key on-the-fly. The product of the current query and the current key is computed after the RoPE.
Next, the DOT operation between the rotated query and the historical key cache takes place, followed by the value projection. This sequencing allows sufficient time for the softmax operation, ensuring the softmax output is prepared before the final stage—summing the values weighted by the softmax probabilities. Finally, the scaled-dot product between the attention score and the values of each historical token is computed.
It is also important to note that the quantization process occurs concurrently as the current key and value are generated.
After all heads complete their computations, the outputs are concatenated and used as input for the output projection operation. As the output projection results are generated, the residual connection is added, and the square sum required for post-attention layer normalization is computed simultaneously. All the miscellaneous operations are hidden in the dense computation of the attention layer, ensuring no cycle penalties.

\subsection{Data Arrangement Format}

\begin{figure}
    \centering
    \includegraphics[width=1.0\linewidth]{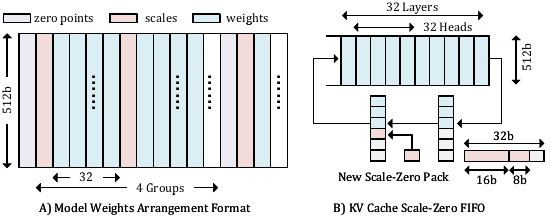}
    \caption{Bus-width Aligned Data Arrangement Format. A) Compact model weight arrangement format interleaving zero points, scales, and weights. B) KV cache scale-zero packing process to minimize scalar data transfers.}
    \label{fig:format}
    \end{figure}

\textbf{One of the most effective strategies for optimizing bandwidth utilization is implementing sequential burst transfers.}
Large consecutive burst transfers can achieve significantly higher bandwidth efficiency compared to short bursts with discontinuous addresses. We propose a customized data arrangement format for model weights and the KV cache to ensure that all transactions occur as large consecutive bursts.

\begin{figure*}
    \centering
    \includegraphics[width=1.0\linewidth]{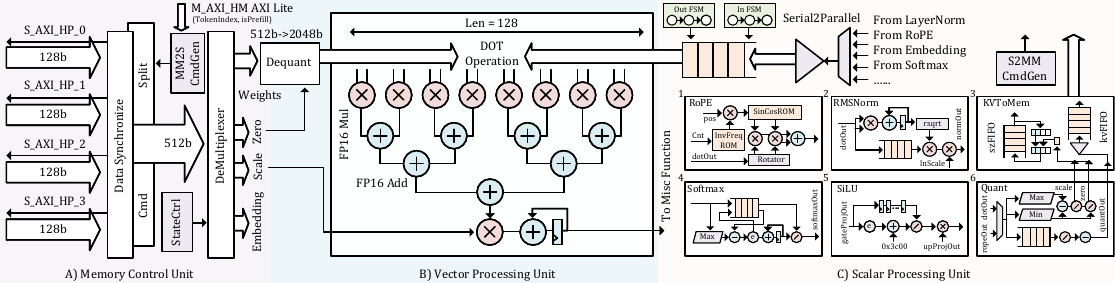}
    \caption{Hardware Architecture of the Accelerator. A) Memory Control Unit (shaded in light purple) ensures full access to DDR bandwidth. B) Vector Processing Unit (shaded in light blue) performs dense computations. C) Scalar Processing Unit (shaded in light orange) handles miscellaneous processes.}
    \label{fig:architecture}
    \end{figure*}

\subsubsection{Model Weight Data Arrangement}

Model quantization generates scales and zero points, which are used to recover the original floating-point values. While fetching the scales and zero points group by group during the quantization process is straightforward, fetching small amounts of data from DDR memory introduces high latency and reduces bandwidth utilization. Alternatively, loading all the scales and zero points at once and storing them in on-chip memory is feasible, but the size of the scales and zero points for a single linear layer is still too large for the local BRAM or URAM.

To address this, we propose a specialized data arrangement format for the 4-bit quantized model weights, where the zero points, scales, and weights are interleaved, as shown in Fig.\ref{fig:format}A. Assuming a 512-bit data bus, each transaction can carry 64 4-bit weights, 64 4-bit zero points, or 16 16-bit scales.
Assuming a quantization group size of 128, each transaction for scales includes all the scales for 2,048 quantized weights, which corresponds to 32 transactions.
Each transaction for zero points contains 4 groups of scales and weights, as illustrated in Fig.\ref{fig:format}A. By packing zero points, scales, and weights together, we ensure consecutive burst transfers, minimizing the on-chip buffer needed to store zero points and scales.
Zero points, scales, and weights are separated for subsequent computation as they flow through the PL logic.

\subsubsection{Key-Value Scale-Zero Packing}

Unlike the scales and zero points for model weights, which can be pre-arranged in a specific layout, the scales and zero points for the KV cache are generated on-the-fly during inference.
To avoid transferring small data volumes for scales and zero points, we maintain a FIFO buffer that stores the packed scales and zero points of the KV cache until element inside the FIFO contains all vaild packs, as shown in Fig.\ref{fig:format}B.
As the inference proceeds head-wise and layer-wise for each token, every time a scale-zero pack is generated, we pop one element from the FIFO, update it by appending the new valid scale-zero pack, and then push the element back into the FIFO. Each scale-zero pack is 32-bit, consisting of 16-bit for the scale, 8-bit for the zero point, and a dummy 8-bit space for data alignment. Assuming a 512-bit data bus, such a FIFO can maintain scale-zero packs for 16 tokens. The scale-zero packs are transferred back to DDR once we begin the inference process for the 16th token.
In this way, we ensure bus-aligned transfer of the KV cache scales and zero points, improving the bandwidth utilization.

\section{Hardware Architecture}

The hardware architecture of the LLM accelerator we proposed consists of three main components: the Memory Control Unit (MCU), the Vector Processing Unit (VPU), and the Scalar Processing Unit (SPU), as shown in Fig.\ref{fig:architecture}.
We will introduce the design of each component in the following subsections.

\subsection{Memory Control Unit}

\textbf{The MCU ensures that the on-chip bandwidth resources are fully accessible to the PL logic.}
The KV260 is equipped with 4GB of 2400 Mbps 64-bit DDR4 on the processing system (PS) side, delivering a bandwidth of 19.2 GB/s. The Zynq UltraScale+ MPSoC provides several AXI ports with a maximum 128-bit, allowing the programmable logic (PL) to access the DDR4.
To align with the available bandwidth of KV260, we utilize four AXI ports, with the system frequency set to 300 MHz, as shown in Fig.\ref{fig:architecture}A.
Four 128-bit streams from the four AXI ports are synchronized, concatenated into a 512-bit stream, and sent to a demultiplexer that separates the scales, zero points, quantized weights, and token embeddings. The processing system (PS) is responsible for tokenizing the input context and sending the token index to the memory command generator via the AXI-Lite bus. The commands are then split into four, one for each AXI port.

\subsection{Vector Processing Unit}

The 512-bit weights are dequantizd to 128 sets of 16-bit floating points data and sent to the VPU for subsequent computation, as shown in Fig.\ref{fig:architecture}B.
End-to-end LLM inference involves both the prefill and decode phases, which have significantly different computing requirements.
In GPU devices, where computing resources are far more abundant than memory resources, the prefill stage involves matrix multiplication, processing multiple tokens at the same time, with most tensor cores operational and on-chip model weights being reused.
In contrast, during the decode stage, most tensor cores are idle due to the bandwidth-bound nature of matrix-vector computations in the "one token at a time" autoregressive decoding process.
However, in domain-specific architecture design where power, performance, and area (PPA) are critical considerations, it is inefficient to have most computing resources idle during the decode stage.
Therefore, we sacrifice some performance in the prefill stage and implement a bandwidth-area balanced DOT computing engine that can fully utilize the bandwidth \textbf{without wasting additional compute resources, reaching the optimized PPA requirements in the decode stage.}
The simple vector dot engine (rather than a matrix engine\cite{li2024revealing}\cite{li2023firefly}) has 128 multipliers, matching the size of the quantized weights set after the dequantization, along with an adder tree for summing the products, a scaling multiplier, and an accumulator to perform the dot operations, shown in Fig.\ref{fig:architecture}B.

It is worth noting that we adopt FP16 computation on FPGA.
Although floating point computation is inherently expensive on FPGAs, the decoding performance bottleneck in LLM inference is the bandwidth, not the logic resources. While recent studies have shown that more aggressive quantization methods can be applied to activations for a fully integer-only inference\cite{hu2024llm}, we aim to maximize the use of existing computing resources by building a floating point compute engine to maintain LLM's accuracy as much as possible.

\subsection{Scalar Processing Unit}

The SPU performs miscellaneous functions concurrently with the VPU to ensure no cycle penalties, as demonstrated in section VA. In this subsection, we briefly introduce the design of the miscellaneous submodules in the SPU as follows.

\textbf{RoPE}.
The RoPE submodule comprises three primary components: the rotator, the sin/cos generator, and the address generator, as shown in Fig.\ref{fig:architecture}C1. The rotator caches half of the query or key and generates the rotation pair. The sin/cos generator includes a look-up table (LUT) that stores 4096 points of one-quarter cycle of sine wave values in read-only memory (ROM). The address generator consists of another LUT that stores the inverted frequency values: $10000.0^{-\frac{i}{4096}},i=0,2,4...4094$, to generate the addresses for reading sine and cosine values based on the current token number. The rotation pair is then multiplied by the sine and cosine values to obtain the rotated query or key.

\textbf{RMSNorm}.
The RMSNorm process involves two sequential passes over the input vector, as shown in Fig.\ref{fig:architecture}C2. The first pass computes the root mean square (RMS) of the input vector: $\mathbf{RMS}(\mathbf{a})=\sqrt{\frac{1}{n}\sum_{i=1}^{n}a_i^2}$; however, this step can be bypassed if the mean square value can be computed by the DOT engine. The second pass performs normalization based on the RMS values obtained: $\overline{a_i}=\frac{a_i}{\mathbf{RMS(a)}}$.

\textbf{Softmax}.
The softmax submodule implements a numerically stable variant\cite{milakov2018online} of the softmax function to mitigate potential numerical issues. This approach involves three sequential passes over the input vector, as shown in Fig.\ref{fig:architecture}C4. The first pass identifies the maximum value $m$ of the input vector. The second pass computes the normalization term $d=\sum_{n}^{i=1}e^{x_i - m}$. Finally, the third pass calculates the resulting softmax values: $s_i = \frac{e^{x_i - m}}{d}$.

\textbf{SiLU}.
The SiLU submodules contains logic pipeline for computing the $\frac{x}{1+e^{-x}}$ function, as shown in Fig.\ref{fig:architecture}C5, where $x$ is the gate projection output in the MLP layer. The SiLU output is then multiplied by the output of the up-projection layer to produce the input for the down-projection layer.

\textbf{Quantization}. 
The quantization submodule operates in two passes over the input vector, as shown in Fig.\ref{fig:architecture}C6. The first pass determines the scaling factor $s=\frac{x_{max}-x_{min}}{255}$ and zero point $z = \lceil \frac{x_{min}}{s}\rceil $, while the second pass performs the quantization of the input: $(x - z)\cdot s$. The generated scale and zero point are temporarily stored in the FIFO mentioned in Section VB, and the quantized output is sent to a serial-to-parallel unit to match the AXI bus width before being transferred back to DDR memory, as shown in Fig.\ref{fig:architecture}C3.

\subsection{On-chip Dataflow of Hidden States}

\textbf{To fully utilize the bandwidth and capacity for model weight transfers, hidden states generated during the inference process are always kept on-chip and not offloaded to DDR.}
The serial outputs from the RMSNorm, RoPE, softmax, and MLP gated outputs are multiplexed and sent to a serial-to-parallel adapter. These hidden states are then pushed to a DOT operand FIFO, where two in-out finite state machines ensures the correct data is sent to the VPU for DOT computation.

\section{Experiments}

\subsection{Basic Setup}

The hardware accelerator is designed using SpinalHDL\cite{spinalhdl}, with behavioral simulations conducted via cocotb\cite{rosser2018cocotb}. The Verilog codes generated by the SpinalHDL compiler are synthesized and implemented using Xilinx Vivado 2022.2. Power consumption estimates are also obtained from the reports generated by the Vivado Design Suite.

The original LLaMA2-7B model is quantized using the AutoAWQ library, converted to our proposed format, loaded onto an SD card, and then transferred to the DDR memory of the KV260 platform in the C bare-metal program.
The 4GB address space in the KV260 platform is divided into two parts: the lower 2GB spans from 0x00000000-0x7FF00000, with 1MB reserved by the compiler, and the higher 2GB spans from 0x80000000-0xFFFFFFFF.
We place the embedding table, model weights, and reserved space for KV cache upto 1024 tokens for the first 16 layers to the higher address space, while the remaining data is placed in the lower address space.
\textbf{It is worth noting that the model weights of the LLaMA2-7B model have reached the limit of the DDR4 memory capacity, making it impossible to load a Linux operating system with so little memory remaining.}

\subsection{Resources Consumption Breakdown}

\begin{table}[]
\centering
\caption{Resources Consumption Breakdown of the Accelerator.}
\label{utilbreakdown}
\begin{tabular}{c|c|c|c|c}
    \hline
          & Total       & MemCtrl     & VPU         & SPU          \\ \hline
          & Used/Util.  & Used/Util.  & Used/Util.  & Used/Util.   \\ \hline\hline
    LUTs  & 78K  / 67\% & 14K  / 12\% & 34K  / 30\% & 29K  / 25\%  \\ \hline
    FFs   & 105K / 45\% & 21K  / 8\%  & 44K  / 19\% & 40K  / 16\%   \\ \hline
    CARRY & 3.8K / 26\% & 0.6K / 4\%  & 2.1K / 14\% & 1K   / 6\%   \\ \hline
    DSP   & 291  / 24\% & 1    / 0    & 266  / 21\% & 24   / 2\%   \\ \hline
    URAM  & 10   / 16\% & 7    / 11\% & 0    / 0    & 3    / 4\%    \\ \hline
    BRAM  & 36.5 / 25\% & 30   / 20\% & 0    / 0    & 6.5  / 5\%   \\ \hline
    \end{tabular}
\end{table}

The resource consumption of the proposed accelerator is shown in Table \ref{utilbreakdown}. The accelerator utilizes 67\% of the LUTs, 45\% of the FFs, 26\% of the CARRYs, 24\% of the DSPs, 16\% of the URAMs, and 25\% of the BRAMs. The VPU consumes the most LUTs and DSPs, as it serves as the core of the accelerator, performing the dense FP16 computations. The MCU consumes the most of the BRAMs and URAMs, as it contains the AXI Datamover IP and buffers the data from the AXI interface.
The power consumption reported by the Vivado is \textbf{6.57W}, with the accelerator operating at \textbf{300MHz}.

\subsection{Comparison with Existing FPGA Research}

\begin{table*}[]
    \centering
    \caption{Performance Comparison with Existing FPGA Research.}
    \label{fpgacompare}
    \resizebox{\textwidth}{!}{
    \begin{threeparttable}
    \begin{tabular}{c|c|c|c|c|c|c|c|c|c|c|c|c|c|c}
    \hline
                                                                         &           & Device  & LUT  & FF    & BRAM  & DSP  & MHz     & W    & GB/s & Tasks          & Opt.     & token/s\tnote{1}   & token/s\tnote{2}  & Util. \%\\ \hline\hline
    \multirow{3}{*}{\begin{tabular}[c]{@{}c@{}}Cloud\\ HBM\end{tabular}} & DFX\tnote{3} & U280    & 520K & 1107K & 1192  & 3533 & 200     & 45   & 460  & GPT2-1.5B      & W16      & $\sim$153 & $\sim$21\tnote{4} & 13.7\%    \\ \cline{2-15} 
                                                                         & FlightLLM & U280    & 574K & 943K  & 1252  & 6345 & 225     & 45   & 460  & LLaMA2-7B      & W4\tnote{5}       & $\sim$131 & $\sim$55 & 42/65.9\%\tnote{6} \\ \cline{2-15} 
                                                                         & EdgeLLM   & U280    & 967K & 607K  & 1734  & 5587 & 250     & 50.7 & 460  & ChatGLM-6B     & W4       & $\sim$153 & $\sim$75 & 49/73.8\%\tnote{7} \\ \hline\hline
    \multirow{3}{*}{\begin{tabular}[c]{@{}c@{}}Edge\\ DDR\end{tabular}}  & SECDA\cite{haris2024designing}     & PYNQ    & /    & /     & /     & /    & /       & /    & 2.1  & TinyLLaMA\tnote{7}      & W4       & 3.8       & 0.58     & 15.2\%    \\ \cline{2-15} 
                                                                         & LlamaF\cite{xu2024llamaf}    & ZCU102  & 164K & 171K  & 223   & 528  & 205     & 5.08 & 21.3 & TinyLLaMA      & W8       & 19.3      & 1.5      & 7.7\%     \\ \cline{2-15} 
                                                                         & Ours      & KV260   & 78K  & 105K  & 36.5  & 291  & 300     & 6.57 & 19.2 & LLaMA2-7B      & W4       & 5.8       & 4.9      & 84.5\%    \\ \hline
    \end{tabular}
    \begin{tablenotes}
        \item[1] The theoretical peak decoding speed of the given bandwidth.
        \item[2] The actual reported decoding performance data.
        \item[3] The table shows the resource utilization of a single FPGA in DFX's designs.
        \item[4] DFX only reports the single FPGA decoding performance for a 345M model. The single FPGA decoding performance for the 1.5B model is extrapolated by linear scaling.
        \item[5] Although FlightLLM uses 8-bit quantization, it employs the SparseGPT method, achieving an effective average bit-width of 3.5 bits, equivalent to 4-bit quantization in terms of capacity and bandwidth. 
        \item[6,7] Both FlightLLM and EdgeLLM report bandwidth utilization metrics higher than theoretical calculations, we list both of the results for reference.
        \item[8] TinyLLaMA model is 1.1B.
    \end{tablenotes}
    \end{threeparttable}
    }
    \end{table*}

Performance comparison between our proposed accelerator and existing research is shown in Table.\ref{fpgacompare}. Bandwidth utilization is a reliable metric for assessing the efficiency of different platforms, as it indicates how closely frameworks can approach the theoretical decoding speed of the hardware backend. Given that decoding speed is entirely bandwidth-bound, utilization is calculated as the ratio of the measured token/s to the number of model weight transfers possible within one second.

Comparisons with research utilizing large cloud FPGAs with HBM\cite{zeng2024flightllm}\cite{hong2022dfx}\cite{huang2024edgellm}\cite{chen2023understanding} can be hard due to the significant bandwidth disparity (460GB/s for Alveo U280 vs. 19.2GB/s for KV260).Nevertheless, for reference, we present the performance of these works in Table.\ref{fpgacompare}.
To the best of our knowledge, little research has been conducted on deploying LLMs on embedded FPGAs. SECDA-LLM \cite{haris2024designing}, evaluates the TinyLLama-1.1B on the Pynq-Z2, achieving a performance of 0.58 token/s. Similarly, LlamaF \cite{xu2024llamaf} deploys the TinyLLama-1.1B on the ZCU102, achieving 1.5 token/s.
While the decoding speed and bandwidth utilization of both are not yet optimal, these works represent some of the earliest efforts to deploy LLMs on edge FPGAs.

\subsection{Comparison with Embedded CPU and GPUs}

Table.\ref{gpucompare} presents a comparison of decoding performance and bandwidth utilization efficiency for the 4-bit quantized LLaMA2-7B inference task between our proposed accelerator on the KV260, the embedded CPU Raspberry Pi, and the Jetson series embedded GPUs, using various inference frameworks.
It is important to note that few frameworks are available for implementing LLMs on edge devices, and these frameworks vary significantly in performance.
The reported decoding performance data is sourced from existing research or official websites of the respective inference frameworks.
The llama.cpp \cite{llamacpp} represents an inference framework that requires minimal setup and delivers good performance across a wide variety of hardware.
However, it results in low token/s and bandwidth utilization on Raspberry Pi or Jetson AGX Orin \cite{dhar2024empirical}.
TinyChatEngine\cite{tinychatengine} is an on-device LLM inference library enabled by LLM model compression and achieves good decoding performance on the Jetson AGX Orin \cite{tinychatperf}. However, the bandwidth utilization remains relatively low.
NanoLLM \cite{nanollm} is a high-performance library with official support for Jetson devices. It achieves excellent decoding performance on Jetson devices, reaching around 80\% bandwidth utilization.
Although the KV260 has significantly less bandwidth compared to the Jetson GPUs, our proposed hardware accelerator achieves superior resource utilization with nearly 85\% bandwidth efficiency, with 6\% higher utilization than the Jetson Orin Nano using the NanoLLM,  as shown in Table.\ref{gpucompare}.

\begin{table}[]
    \centering
    \caption{Comparison with Embeded GPUs in 4-bit LLaMA2-7B inference}
    \label{gpucompare}
    \begin{threeparttable}
    \begin{tabular}{c|c|c|c|c|c}
    \hline
    Device                           & GB/s                   & FrameWork & token/s\tnote{1} & token/s\tnote{2}    & Util.      \\ \hline\hline
    Pi-4B 8GB                        & 12.8                   & llama.cpp & 3.9     & 0.11\cite{dhar2024empirical} & 2.8\%      \\ \hline
    \multirow{3}{*}{JetsonAGXOrin}   & \multirow{3}{*}{204.8} & llama.cpp & 62.5    & 4.49\cite{dhar2024empirical} & 7.2\%      \\ \cline{3-5} 
                                     &                        & TinyChat  & 62.5    & 33\cite{tinychatperf}        & 52.8\%     \\ \cline{3-5} 
                                     &                        & NanoLLM   & 62.5    & 47.1\cite{jetsonbenchmark}   & 75.4\%     \\ \hline
    JetsonOrinNano                   & 68                     & NanoLLM   & 20.7    & 16.4\cite{jetsonbenchmark}   & 79.2\%     \\ \hline
    KV260                            & 19.2                   & Ours      & 5.8     & 4.9                          & 84.5\%     \\ \hline
    \end{tabular}
    \begin{tablenotes}
        \item[1,2] The theoretical peak decoding speed and the actual reported speed.
      \end{tablenotes}
    \end{threeparttable}
    \end{table}

\section{Discussion and Conclusion}

We present several insights for designing architectures for LLM inference with embedded FPGAs in this section.

\textbf{Memory Resources is Essential.}
The memory capacity of an device determines the feasibility of LLM deployment, and the performance of LLM decoding is directly tied to the available bandwidth. We have pushed up to the limit by deploying a 7B model on a 4GB device in this work.
\textbf{Nevertheless, further improving LLM decoding speed and supporting larger LLM size remains challenging without sufficient bandwidth and capacity.}
With DDR5 and unified memory—both of which offer increased bandwidth—becoming more common in laptops and smartphones, it is timely for FPGA vendors to integrate advanced memory support into embedded devices.

\textbf{Bandwidth Utilization is Critical.} When bandwidth is limited, it is essential to maximize the utilization. For CPU or GPU hardware, precisely controlling on-chip cache behavior to reserve full bandwidth for non-temporal weight transfers can be hard. This is where specialized architectures with customized cache and dataflow offer a significant advantage.

In this work, we introduce an LLM accelerator based on embedded FPGAs, capable of supporting models with up to 7 billion parameters by maximizing bandwidth and capacity utilization. As the first research to deploy a 7B LLM on an embedded FPGA, this study offers insights for the domain-specific architecture community in LLM inference applications and underscores the potential of FPGA-based LLM solutions.

\section{Acknowledgement}


This work, as part of the software-hardware codesigns research of the BrainCog Engine\cite{braincogweb}\cite{Zeng2023}, is supported by the Chinese Academy of Sciences Foundation Frontier Scientific Research Program (ZDBS-LY- JSC013) and a funding from Institute of Automation, Chinese Academy of Sciences (Grant No. E411230101).
We would also like to thank Hongzheng Zhang and Niansong Zhang from Cornell University for providing statistics of the paper\cite{chen2023understanding}.

\bibliographystyle{IEEEtran}
\bibliography{myIEEE,reference}

\begin{thebibliography}{10}
\providecommand{\url}[1]{#1}
\csname url@samestyle\endcsname
\providecommand{\newblock}{\relax}
\providecommand{\bibinfo}[2]{#2}
\providecommand{\BIBentrySTDinterwordspacing}{\spaceskip=0pt\relax}
\providecommand{\BIBentryALTinterwordstretchfactor}{4}
\providecommand{\BIBentryALTinterwordspacing}{\spaceskip=\fontdimen2\font plus
\BIBentryALTinterwordstretchfactor\fontdimen3\font minus \fontdimen4\font\relax}
\providecommand{\BIBforeignlanguage}[2]{{%
\expandafter\ifx\csname l@#1\endcsname\relax
\typeout{** WARNING: IEEEtran.bst: No hyphenation pattern has been}%
\typeout{** loaded for the language `#1'. Using the pattern for}%
\typeout{** the default language instead.}%
\else
\language=\csname l@#1\endcsname
\fi
#2}}
\providecommand{\BIBdecl}{\relax}
\BIBdecl

\bibitem{guo2017angel}
Kaiyuan Guo, Lingzhi Sui, Jiantao Qiu, Jincheng Yu, Junbin Wang, Song Yao, Song Han, Yu~Wang, and Huazhong Yang, ``Angel-eye: A complete design flow for mapping cnn onto embedded fpga,'' \emph{IEEE transactions on computer-aided design of integrated circuits and systems}, vol.~37, no.~1, pp. 35--47, 2017.

\bibitem{chen2021hardware}
Xiang Chen, Jindong Li, and Yong Zhao, ``Hardware resource and computational density efficient cnn accelerator design based on fpga,'' in \emph{2021 IEEE International Conference on Integrated Circuits, Technologies and Applications (ICTA)}.\hskip 1em plus 0.5em minus 0.4em\relax IEEE, 2021, pp. 204--205.

\bibitem{li2024firefly}
Jindong Li, Guobin Shen, Dongcheng Zhao, Qian Zhang, and Yi~Zeng, ``Firefly v2: Advancing hardware support for high-performance spiking neural network with a spatiotemporal fpga accelerator,'' \emph{IEEE Transactions on Computer-Aided Design of Integrated Circuits and Systems}, 2024.

\bibitem{li2024fireflys}
Tenglong Li, Jindong Li, Guobin Shen, Dongcheng Zhao, Qian Zhang, and Yi~Zeng, ``Firefly-s: Exploiting dual-side sparsity for spiking neural networks acceleration with reconfigurable spatial architecture,'' \emph{IEEE Transactions on Circuits and Systems I: Regular Papers}, 2024.

\bibitem{li2022auto}
Zhengang Li, Mengshu Sun, Alec Lu, Haoyu Ma, Geng Yuan, Yanyue Xie, Hao Tang, Yanyu Li, Miriam Leeser, Zhangyang Wang \emph{et~al.}, ``Auto-vit-acc: An fpga-aware automatic acceleration framework for vision transformer with mixed-scheme quantization,'' in \emph{2022 32nd International Conference on Field-Programmable Logic and Applications (FPL)}.\hskip 1em plus 0.5em minus 0.4em\relax IEEE, 2022, pp. 109--116.

\bibitem{dong2023heatvit}
Peiyan Dong, Mengshu Sun, Alec Lu, Yanyue Xie, Kenneth Liu, Zhenglun Kong, Xin Meng, Zhengang Li, Xue Lin, Zhenman Fang \emph{et~al.}, ``Heatvit: Hardware-efficient adaptive token pruning for vision transformers,'' in \emph{2023 IEEE International Symposium on High-Performance Computer Architecture (HPCA)}.\hskip 1em plus 0.5em minus 0.4em\relax IEEE, 2023, pp. 442--455.

\bibitem{hong2022dfx}
Seongmin Hong, Seungjae Moon, Junsoo Kim, Sungjae Lee, Minsub Kim, Dongsoo Lee, and Joo-Young Kim, ``Dfx: A low-latency multi-fpga appliance for accelerating transformer-based text generation,'' in \emph{2022 55th IEEE/ACM International Symposium on Microarchitecture (MICRO)}.\hskip 1em plus 0.5em minus 0.4em\relax IEEE, 2022, pp. 616--630.

\bibitem{zeng2024flightllm}
Shulin Zeng, Jun Liu, Guohao Dai, Xinhao Yang, Tianyu Fu, Hongyi Wang, Wenheng Ma, Hanbo Sun, Shiyao Li, Zixiao Huang \emph{et~al.}, ``Flightllm: Efficient large language model inference with a complete mapping flow on fpga,'' \emph{arXiv preprint arXiv:2401.03868}, 2024.

\bibitem{touvron2023llama}
Hugo Touvron, Louis Martin, Kevin Stone, Peter Albert, Amjad Almahairi, Yasmine Babaei, Nikolay Bashlykov, Soumya Batra, Prajjwal Bhargava, Shruti Bhosale \emph{et~al.}, ``Llama 2: Open foundation and fine-tuned chat models,'' \emph{arXiv preprint arXiv:2307.09288}, 2023.

\bibitem{zhang2019root}
Biao Zhang and Rico Sennrich, ``Root mean square layer normalization,'' \emph{Advances in Neural Information Processing Systems}, vol.~32, 2019.

\bibitem{milakov2018online}
Maxim Milakov and Natalia Gimelshein, ``Online normalizer calculation for softmax,'' \emph{arXiv preprint arXiv:1805.02867}, 2018.

\bibitem{su2024roformer}
Jianlin Su, Murtadha Ahmed, Yu~Lu, Shengfeng Pan, Wen Bo, and Yunfeng Liu, ``Roformer: Enhanced transformer with rotary position embedding,'' \emph{Neurocomputing}, vol. 568, p. 127063, 2024.

\bibitem{elfwing2018sigmoid}
Stefan Elfwing, Eiji Uchibe, and Kenji Doya, ``Sigmoid-weighted linear units for neural network function approximation in reinforcement learning,'' \emph{Neural networks}, vol. 107, pp. 3--11, 2018.

\bibitem{vaswani2017attention}
A~Vaswani, ``Attention is all you need,'' \emph{Advances in Neural Information Processing Systems}, 2017.

\bibitem{dosovitskiy2020image}
Alexey Dosovitskiy, ``An image is worth 16x16 words: Transformers for image recognition at scale,'' \emph{arXiv preprint arXiv:2010.11929}, 2020.

\bibitem{devlin2018bert}
Jacob Devlin, Ming-Wei Chang, Kenton Lee, and Kristina Toutanova, ``Bert: Pre-training of deep bidirectional transformers for language understanding,'' \emph{arXiv preprint arXiv:1810.04805}, 2018.

\bibitem{liu2021hardware}
Zejian Liu, Gang Li, and Jian Cheng, ``Hardware acceleration of fully quantized bert for efficient natural language processing,'' in \emph{2021 Design, Automation \& Test in Europe Conference \& Exhibition (DATE)}.\hskip 1em plus 0.5em minus 0.4em\relax IEEE, 2021, pp. 513--516.

\bibitem{radford2019language}
Alec Radford, Jeffrey Wu, Rewon Child, David Luan, Dario Amodei, Ilya Sutskever \emph{et~al.}, ``Language models are unsupervised multitask learners,'' \emph{OpenAI blog}, vol.~1, no.~8, p.~9, 2019.

\bibitem{chen2023understanding}
Hongzheng Chen, Jiahao Zhang, Yixiao Du, Shaojie Xiang, Zichao Yue, Niansong Zhang, Yaohui Cai, and Zhiru Zhang, ``Understanding the potential of fpga-based spatial acceleration for large language model inference,'' \emph{arXiv preprint arXiv:2312.15159}, 2023.

\bibitem{huang2024edgellm}
Mingqiang Huang, Ao~Shen, Kai Li, Haoxiang Peng, Boyu Li, and Hao Yu, ``Edgellm: A highly efficient cpu-fpga heterogeneous edge accelerator for large language models,'' \emph{arXiv preprint arXiv:2407.21325}, 2024.

\bibitem{xiao2023smoothquant}
Guangxuan Xiao, Ji~Lin, Mickael Seznec, Hao Wu, Julien Demouth, and Song Han, ``Smoothquant: Accurate and efficient post-training quantization for large language models,'' in \emph{International Conference on Machine Learning}.\hskip 1em plus 0.5em minus 0.4em\relax PMLR, 2023, pp. 38\,087--38\,099.

\bibitem{lin2024awq}
Ji~Lin, Jiaming Tang, Haotian Tang, Shang Yang, Wei-Ming Chen, Wei-Chen Wang, Guangxuan Xiao, Xingyu Dang, Chuang Gan, and Song Han, ``Awq: Activation-aware weight quantization for on-device llm compression and acceleration,'' \emph{Proceedings of Machine Learning and Systems}, vol.~6, pp. 87--100, 2024.

\bibitem{li2024evaluating}
Shiyao Li, Xuefei Ning, Luning Wang, Tengxuan Liu, Xiangsheng Shi, Shengen Yan, Guohao Dai, Huazhong Yang, and Yu~Wang, ``Evaluating quantized large language models,'' \emph{arXiv preprint arXiv:2402.18158}, 2024.

\bibitem{li2024revealing}
Jindong Li, Tenglong Li, Guobin Shen, Dongcheng Zhao, Qian Zhang, and Yi~Zeng, ``Revealing untapped dsp optimization potentials for fpga-based systolic matrix engines,'' in \emph{2024 34th International Conference on Field-Programmable Logic and Applications (FPL)}.\hskip 1em plus 0.5em minus 0.4em\relax IEEE, 2024, pp. 197--203.

\bibitem{li2023firefly}
Jindong Li, Guobin Shen, Dongcheng Zhao, Qian Zhang, and Yi~Zeng, ``Firefly: A high-throughput hardware accelerator for spiking neural networks with efficient dsp and memory optimization,'' \emph{IEEE Transactions on Very Large Scale Integration (VLSI) Systems}, vol.~31, no.~8, pp. 1178--1191, 2023.

\bibitem{hu2024llm}
Xing Hu, Yuan Chen, Dawei Yang, Sifan Zhou, Zhihang Yuan, Jiangyong Yu, and Chen Xu, ``I-llm: Efficient integer-only inference for fully-quantized low-bit large language models,'' \emph{arXiv preprint arXiv:2405.17849}, 2024.

\bibitem{spinalhdl}
\BIBentryALTinterwordspacing
``Spinalhdl: Scala based hdl.'' [Online]. Available: \url{https://github.com/SpinalHDL/SpinalHDL}
\BIBentrySTDinterwordspacing

\bibitem{rosser2018cocotb}
Benjamin~John Rosser, ``Cocotb: a python-based digital logic verification framework,'' in \emph{Micro-electronics Section seminar. CERN, Geneva, Switzerland}, 2018.

\bibitem{haris2024designing}
Jude Haris, Rappy Saha, Wenhao Hu, and Jos{\'e} Cano, ``Designing efficient llm accelerators for edge devices,'' \emph{arXiv preprint arXiv:2408.00462}, 2024.

\bibitem{xu2024llamaf}
\BIBentryALTinterwordspacing
Han Xu, Yutong Li, and Shihao Ji, ``Llamaf: An efficient llama2 architecture accelerator on embedded fpgas,'' 2024. [Online]. Available: \url{https://arxiv.org/abs/2409.11424}
\BIBentrySTDinterwordspacing

\bibitem{llamacpp}
\BIBentryALTinterwordspacing
``Georgi gerganov. ggerganov/llama.cpp: Port of facebook’s llama model in c/c++.'' [Online]. Available: \url{https://github.com/ggerganov/}
\BIBentrySTDinterwordspacing

\bibitem{dhar2024empirical}
Nobel Dhar, Bobin Deng, Dan Lo, Xiaofeng Wu, Liang Zhao, and Kun Suo, ``An empirical analysis and resource footprint study of deploying large language models on edge devices,'' in \emph{Proceedings of the 2024 ACM Southeast Conference}, 2024, pp. 69--76.

\bibitem{tinychatengine}
\BIBentryALTinterwordspacing
``Tinychat: Large language model on the edge.'' [Online]. Available: \url{https://hanlab.mit.edu/blog/tinychat}
\BIBentrySTDinterwordspacing

\bibitem{tinychatperf}
\BIBentryALTinterwordspacing
``Tinychat: Efficient and lightweight chatbot with awq.'' [Online]. Available: \url{https://github.com/mit-han-lab/llm-awq/blob/main/tinychat/README.md}
\BIBentrySTDinterwordspacing

\bibitem{nanollm}
\BIBentryALTinterwordspacing
``Optimized local inference for llms with huggingface-like apis for quantization, vision/language models, multimodal agents, speech, vector db, and rag.'' [Online]. Available: \url{https://dusty-nv.github.io/NanoLLM/}
\BIBentrySTDinterwordspacing

\bibitem{jetsonbenchmark}
\BIBentryALTinterwordspacing
``Jetson ai lab benchmark.'' [Online]. Available: \url{https://www.jetson-ai-lab.com/benchmarks.html}
\BIBentrySTDinterwordspacing

\bibitem{braincogweb}
\BIBentryALTinterwordspacing
``Braincog: Brain-inspired cognitive intelligence engine.'' [Online]. Available: \url{http://www.brain-cog.network}
\BIBentrySTDinterwordspacing

\bibitem{Zeng2023}
\BIBentryALTinterwordspacing
Yi~Zeng, Dongcheng Zhao, Feifei Zhao, Guobin Shen, Yiting Dong, Enmeng Lu, Qian Zhang, Yinqian Sun, Qian Liang, Yuxuan Zhao, Zhuoya Zhao, Hongjian Fang, Yuwei Wang, Yang Li, Xin Liu, Chengcheng Du, Qingqun Kong, Zizhe Ruan, and Weida Bi, ``{BrainCog}: A spiking neural network based, brain-inspired cognitive intelligence engine for brain-inspired {AI} and brain simulation,'' \emph{Patterns}, p. 100789, Jul. 2023. [Online]. Available: \url{https://doi.org/10.1016/j.patter.2023.100789}
\BIBentrySTDinterwordspacing

\end{thebibliography}

\end{document}